\documentclass{article}
\usepackage[margin=1in]{geometry}
\usepackage{amsmath}
\usepackage{graphicx}
\usepackage{float}
\usepackage{subcaption}
\usepackage{hyperref}
\usepackage{lineno}
\usepackage[numbers]{natbib}

\begin{document}

\title{On quantifying the spin angular momentum density of light}
\author{Xiaoyu Zheng, Peter Palffy-Muhoray \\
Department of Mathematical Sciences and \\
Advanced Materials and Liquid Crystal Institute,\\
Kent State University, Kent, OH}
\maketitle
\date{}

\begin{abstract}
In addition to energy, light carries linear and angular momentum. These are
key quantities in rapidly developing optics research and in technologies
focusing on light induced forces and torques on materials. Spin angular
momentum (SAM) density is of particular interest, since unlike orbital
angular momentum, it is uncoupled from linear momentum. The SAM density of
light was first estimated in 1909 by Poynting, using a mechanical analogy.
Exact expressions, based on results from quantum mechanics and field theory
were subsequently developed, and are in common use today. In this paper, we
show that the SAM density of light can be obtained directly from the Coulomb
force and Maxwell's equations, without reliance on quantum mechanics or
field theories; it could have been calculated by Maxwell and his
contemporaries. Besides its historical significance, the simple derivation
of our result makes it readily accessible to non-experts in the field.
\end{abstract}

\section{Introduction}

Although light consists of massless photons, it carries not only energy, but
also linear and angular momentum. Its subsequent ability to exert forces and
torques has opened the door to a fascinating world of optomechanical
phenomena. Great advances have been made in recent years in the fundamental
understanding of light matter interactions, particularly in the areas of
optical torques and angular momentum transport. The angular momentum of
light is traditionally separated into spin and orbital contributions. In
this paper we focus on optical torque and spin angular momentum (SAM) in the
simple case of plane waves.

Plane waves, where the fields have no spatial variation in the plane normal
to the direction of propagation, are infinite in extent and hence do not
exist in nature any more than, say, Gaussian beams. Nonetheless, plane waves
have been useful in the past in providing insights into optical phenomena in
regions of space where the existing real fields resemble plane waves. We
focus on plane waves with this perspective in this work.

Johannes Kepler (1571-1630), on observing that comet tails point away from
the Sun, proposed that light carries linear momentum. Maxwell (1831-1879)
was well aware of the existence of linear momentum carried by light and was
able to calculate radiation pressure on a mirror \cite{maxwell1873treatise}.
There is no evidence, however, that Maxwell was aware of angular momentum
carried by light. In 1905, Einstein \cite{einstein1905erzeugung},
considering the photoelectric effect, argued that light is quantized and
photons have energy $h\nu$, where $h$ is Planck's constant and $\nu $ is the
frequency. In 1909, Poynting \cite{poynting1909wave} proposed that light
carries angular momentum as well as linear momentum, and, noting that the
ratio of angular momentum to energy has units of time, proposed that a
photon has SAM $\mathbf{\pm}\hbar$. Compton's experiments \cite%
{compton1923spectrum} in 1923 confirmed that the linear momentum of a photon
is $h/\lambda$. Orbital angular momentum was discovered in 1992 by Allen 
\textit{et al.}~\cite{allen1992orbital}, who observed that the orbital
angular momentum of a photon is quantized, with values $\pm m\hbar$ where $m 
$ is an integer.

\subsection{The SAM density}

In 1932, C.G. Darwin \cite{darwin1932notes} defined the total angular
momentum of a wave packet of light as%
\begin{equation}
\mathbf{J=}\frac{1}{c^{2}}\int \mathbf{r\times (E\times H)}dV,
\label{Darwin}
\end{equation}%
and, using Maxwell's equations and performing integration by parts,
separated the expression for angular momentum $\mathbf{J}$ into two terms.
The first `represents $\mathbf{r\times P}$ which is the angular momentum of
a particle of momentum $\mathbf{P}$' and the second term, 
\begin{equation}
\rho _{s}=\frac{\varepsilon _{0}}{2\omega }\text{Im}\left( \mathbf{E}^{\ast
}\times \mathbf{E}\right) ,  \label{D2}
\end{equation}%
representing a quantity `analogous to the spin of the electron'. To our
knowledge, this is the first quantification of the SAM density of light in
terms of the Maxwell fields. (In the above expressions, we have altered
notation to enable comparison with recent literature.)

Using essentially the same approach, an expression equivalent to Eq.~%
\eqref{D2} for the SAM density of light was proposed in 1971 by Izmest'ev%
\footnote{%
A.A. Izmest'ev, Classical theory of wave beams, Sov. Phys. J. 14 (1971)
77--80. Translated from A.A. Izmestev, Izvestiya Vysshikh. Uchebnykh
Zavedenii Fizika 1, 101--105 (1971).}, in 1994 by Barnett and Allan \cite%
{barnett1994orbital}, and in 1998 by Berry \cite{berry1998paraxial}. In
2009, using a different but related approach, Berry \cite{berry2009optical}
proposed a decomposition of the Poynting vector and arrived at Eq.~\eqref{D2}%
. In the above derivations, integration by parts was utilized, requiring
that the fields involved vanish at infinity. In addition, in each of the
above arguments, the physical interpretation of the two terms as
representing orbital and SAM is postulated, but not proved; Ref.~\cite%
{barnett1994orbital} cautions about the ready interpretation of these terms.
After Berry's 2009 paper, the expression in Eq.~\eqref{D2} for the SAM
density of light became widely accepted; see for example \cite%
{vernon2024decomposition}.

In light of the considerable effort of researchers in obtaining Eq.~%
\eqref{D2}, it is interesting to inquire about its validity. The logic of
obtaining the angular momentum of light via Eq.~\eqref{Darwin} is predicated
on the classical analogy: the Poynting vector divided by the speed of light
gives the linear momentum density, and the classical angular momentum is the
moment of the linear momentum density. Does the classical analogy hold? The
answer is no. The Poynting vector for right circularly polarized light is
the same as for left circularly polarized light; the SAM density cannot be
obtained from the Poynting vector alone. Nonetheless, remarkably, Eq.~%
\eqref{D2} is valid, as can be readily shown. The expression for the gauge
dependent canonical SAM density of the electromagnetic field is%
\begin{equation}
\mathbf{\rho }_{s\_can}=\varepsilon _{0}\mathbf{E\times A},
\end{equation}%
where $\mathbf{A}$ is the vector potential, an expression first derived by
Belinfante \cite{belinfante1940current} in 1940 for neutrinos. Since $%
\mathbf{E}=-\partial _{t}\mathbf{A}$ for plane waves, after time averaging,
this reduces to the gauge independent Eq.~\eqref{D2}. Details of the formal
derivation of Eq.~\eqref{D2} are given by Bliokh \cite%
{bliokh2014conservation}. We note that the canonical Noether's theorem
approach to the formal derivation and proof also requires that the fields
vanish at infinity.

An alternate empirical proof is suggested by Feynman \cite%
{feynman1963feynman} \textit{et al.} Noting that left- and right-circularly
polarized plane waves are orthogonal eigenfunctions of the wave equation,
elliptically polarized light can be expressed as linear combinations of
these modes. Since the photon density in the modes is given by the normal
mode amplitudes, the SAM density can be calculated at once if the photon
spin is known, giving Eq.~\eqref{D2}. The fields in this approach need not
vanish.

Since Eq.~\eqref{D2} is correct, one is compelled to ask: how is it possible
that the Poynting vector, which carries incomplete information about spin,
can be used to determine the SAM density? Our answer is that in the
derivations used, in addition to the Poynting vector, Maxwell's equations
are relied on as well; additional information, which allows the SAM density
to be calculated, comes from Maxwell's equations. We argue here that,
together with Coulomb's law, Maxwell's equations can give the SAM density,
without the need for the Poynting vector, quantum mechanics, or field
theory. Demonstrating this is the main point of our work.

\subsection{The plane wave paradox}

In 1936, Beth's landmark experiment \cite{beth1936mechanical} showed that a
normally incident circularly polarized light wave, resembling a truncated
plane wave, exerts a torque on a waveplate along the normal. (We distinguish
plane waves with infinite extent and finite aperture or truncated plane
waves, which resemble plane waves in some limited region of space.) In 1954,
Heitler \cite{heitler1954quantum} indicated that according to Eq.~%
\eqref{Darwin}, a plane wave can have no angular momentum in the direction
of propagation. Since a plane wave has no position, it cannot have position
dependent orbital angular momentum, and so it then cannot have spin in the
propagation direction.\footnote{%
For paraxial waves, such as plane waves, spin is along the propagation
direction \cite{bliokh2015transverse}.} This is in apparent contradiction to
the results of Beth's experiment, and gave rise to considerable discussion
in the literature - see, for example \cite{stewart2005angular}.

On one hand, textbooks \cite{feynman1963feynman}, \cite%
{saleh2019fundamentals}, claim that plane waves carry angular momentum,
while erudite papers \cite{allen2002response} \cite{barnett2016natures}
argue that they do not.

One resolution, offered by Stewart \cite{stewart2005angular} along the lines
proposed by Heitler \cite{heitler1954quantum}, is that for a wave of finite
extent, such as Beth's truncated plane wave, the fields at the boundary of
the planar region will generate angular momentum along the propagation
direction \cite{allen2002response, khrapko2001question, yurchenko2002answer}%
. This view was widely adopted; for example, in J.D. Jackson's Classical
Electrodynamics \cite{jackson1999classical}, on page 350, in problem 7.28,
the reader is asked to show that a circularly polarized wave with finite
extent in $x-$ and $y-$ directions, possesses field components along $z$,
and carries angular momentum in this direction.

Subsequent recognition that distinct formalisms and definitions exist for
kinetic (Poynting type) and canonical (Noether's theorem based) momenta with
essential agreement on observables helped resolve conflicts and ambiguities.
An extensive and thorough overview is provided by Bliokh \textit{et al.} \cite%
{bliokh2017optical}.

It appears, however, that in spite of the above developments, the question
of the SAM density of plane waves is not fully resolved. Heitler \cite%
{heitler1954quantum} and adherents argue that plane waves do not carry spin;
recent papers \cite{vernon2024decomposition} still talk about `virtual' SAM
momentum. The arguments and the rigorous proof of Eq.~\eqref{D2} do not hold
for plane waves due to the requirement of fields vanishing at infinity; only
the empirical argument of Feynman \textit{et al.}, predicated on results of
quantum mechanics, does. Using Coulomb's law and Maxwell's equations, we
show that elliptically polarized plane waves do carry SAM. This is the
second point of our work reported in this paper.

Our work is described below.

\section{The SAM current density of plane waves}

We consider the torque exerted by a normally incident elliptically polarized
plane wave on a waveplate. Our approach is purely classical, using only the
Lorentz force and Maxwell's equations. Such a microscopic approach \cite%
{zheng2015electrical} has proved useful in the past. To illustrate the
validity of our approach, we first calculate the radiation pressure - stress
- on an isotropic lossless slab exerted by a normally incident linearly polarized
plane wave. The resulting expression for the stress in terms of the external
fields gives the linear momentum current density of light, indicating the
viability of this approach. We next calculate the areal torque density -
couple stress - on a waveplate exerted by a normally incident elliptically
polarized plane wave. The resulting expression for the couple stress in
terms of the external fields gives the SAM density of light. This is our key
result.

For simplicity and ready accessibility, we use the full time-dependent
expressions for the fields. We include internal reflections in our model,
without which our results would not hold.

\subsection{Radiation Pressure}

We consider an illuminated isotropic lossless slab in vacuum, infinite in the $x$ and 
$y$ directions, with thickness $d$ in the $z$ direction. The electric field
of the incident light has the form%
\begin{equation}
\mathbf{E}_{i}=E_{ix0}\cos (kz-\omega t+\delta )\mathbf{\hat{x}.}
\end{equation}%
In addition to the incident field $\mathbf{E}_{i}$, there are the fields $%
\mathbf{E}_{r}$ and $\mathbf{E}_{t}$ representing reflected and transmitted light
outside the sample, and $\mathbf{E}_{f}$ and $\mathbf{E}_{b}$ representing forward
and backward propagating light inside the sample, with similar form. There
are also corresponding polarization $\mathbf{P}$ and magnetic $\mathbf{H}$
and $\mathbf{B}$ fields.

The radiation pressure, that is, the normal force per area $\mathbf{F}_{A}$
on the slab, can be obtained from the Lorentz force on charges inside the
sample due to the macroscopic Maxwell fields. For lossless and nonmagnetic
materials, this is 
\begin{equation}
\mathbf{F}_{A}=\int_{0}^{d}\langle \mathbf{\dot{P}\times B}\rangle dz,
\end{equation}%
where $\mathbf{P=P}_{f}+\mathbf{P}_{b}$, $\mathbf{B=B}_{f}+\mathbf{B}_{b}$,
and the angle brackets $\langle \rangle $ indicate time average.
Straightforward calculations (see Appendix A), essentially expressing the
fields inside the material in terms of those outside, give the identity%
\begin{equation}
\mathbf{F}_{A}=\frac{1}{c}\langle \mathbf{E}_{i}\times \mathbf{H}_{i}\rangle
-\frac{1}{c}\langle \mathbf{E}_{r}\times \mathbf{H}_{r}\rangle -\frac{1}{c}%
\langle \mathbf{E}_{t}\mathbf{\times H}_{t}\rangle ,  \label{F1}
\end{equation}%
which gives the linear momentum current density of plane waves in vacuum.
This is our first result, which follows directly from Maxwell's equations
and the Lorentz force without recourse to quantum mechanics or field theory.

Our Eq.~\eqref{F1} demonstrates that a linearly polarized plane wave has an
associated tensor linear momentum current density, 
\begin{equation}
\boldsymbol{\varphi }_{L}=\frac{1}{c}\langle \mathbf{E}\times \mathbf{H}%
\rangle \mathbf{\hat{k}},
\end{equation}%
and a vector linear momentum density 
\begin{equation}
\boldsymbol{\rho }_{L}=\frac{1}{c}\boldsymbol{\varphi }_{L}\cdot \mathbf{\ 
\hat{k}}=\frac{1}{c^{2}}\langle \mathbf{E}\times \mathbf{H}\rangle .
\end{equation}

Our derivation of this well known result was included to demonstrate the
effectiveness of our approach. It is interesting to note that there is a
position dependent body force everywhere inside the sample, whose integral
gives the radiation pressure. We note that without internal reflection, the
time averaged body force vanishes; a semi-infinite slab without internal
reflection feels no radiation pressure \cite{frias2012electromagnetic}.

\subsection{Torque on a Waveplate}

We next turn to the problem of our main interest; the optical torque on a
waveplate. We consider a uniaxial waveplate in vacuum, infinite in the $x$
and $y$ directions, with thickness $d$ in the $z$ direction; its optic axis
is along the $x$ direction. The electric field of the incident light has the
form%
\begin{equation}
\mathbf{E}_{i}=E_{ix0}\cos (kz-\omega t+\delta _{x})\mathbf{\hat{x}+}%
E_{iy0}\cos (kz-\omega t+\delta _{y})\mathbf{\hat{y}}.
\end{equation}%
In addition to $\mathbf{E}_{i}$, there are fields $\mathbf{E}_{r}$ and $%
\mathbf{E}_{t}$ representing reflected and transmitted light outside the sample, and 
$\mathbf{E}_{f}$ and $\mathbf{E}_{b}$ representing forward and backward propagating
light inside the sample, with similar form.

We calculate the areal torque density $\mathbf{\tau }_{A}$ on the waveplate
from the Coulomb force on charges inside the sample due to the Maxwell
fields. For nonmagnetic lossless dielectric materials, this is

\begin{equation}
\boldsymbol{\tau }_{A}=\int_{0}^{d}\langle \mathbf{P}\times \mathbf{E}%
\rangle dz,
\end{equation}%
where $\mathbf{E}=\mathbf{E}_{f}+\mathbf{E}_{b}$. We note that Beth \cite%
{beth1936mechanical} expressed light induced torque the same way, but
considered the effects only of the forward propagating wave. As indicated in
the previous section, internal reflections are needed for force and torque
balance in a finite slab with finite thickness.

Straightforward but lengthy calculations (see Appendix B), essentially
expressing the fields inside the material in terms of those outside, give
the identity 
\begin{equation}
\boldsymbol{\tau }_{A}=\frac{\varepsilon _{0}c}{\omega ^{2}}(\mathbf{\dot{E}}%
_{i}\times \mathbf{E}_{i})-\frac{\varepsilon _{0}c}{\omega ^{2}}(\mathbf{%
\dot{E}}_{r}\times \mathbf{E}_{r})-\frac{\varepsilon _{0}c}{\omega ^{2}}(%
\mathbf{\dot{E}}_{t}\times \mathbf{E}_{t}),  \label{ID}
\end{equation}%
which gives the angular momentum current density for plane waves in vacuum.
This is our main result, which again follows solely from Maxwell's equations
and the Lorentz (Coulomb) force. The angular momentum here is spin, since
plane waves do not carry orbital angular momentum.

This identity demonstrates that elliptically polarized plane waves in vacuum
have an associated pseudotensor SAM current density, 
\begin{equation}
\boldsymbol{\varphi }_{s}=\frac{\varepsilon _{0}c}{\omega ^{2}}(\mathbf{\dot{%
E}\times E)\hat{k}},  \label{I1}
\end{equation}%
and a pseudovector SAM density 
\begin{equation}
\boldsymbol{\rho}_{s}=\frac{1}{c}\boldsymbol{\varphi }_{s}\cdot \mathbf{\hat{%
k}}=\frac{\varepsilon _{0}}{\omega ^{2}}(\mathbf{\dot{E}}\times \mathbf{E}).
\label{I2}
\end{equation}

In phasor representation, we have the SAM current density as%
\begin{equation}
\boldsymbol{\varphi }_{s}=\frac{\varepsilon _{0}c}{2\omega }\text{Im}(%
\mathbf{E}^{\ast }\times \mathbf{E)\hat{k}},
\end{equation}%
and the SAM density as 
\begin{equation}
\boldsymbol{\rho }_{s}=\frac{1}{c}\boldsymbol{\varphi }_{s}\cdot \mathbf{%
\hat{k}}=\frac{\varepsilon _{0}}{2\omega }\text{Im}(\mathbf{E}^{\ast }\times 
\mathbf{E}),
\end{equation}%
in agreement with Eq.~\eqref{D2}.

\section{Summary}

Our main result, the identity of areal torque density and the net light SAM
current density, indicates that elliptically polarized plane waves indeed
carry SAM, with a SAM current density 
\begin{equation}
\boldsymbol{\varphi }_{s}=\frac{\varepsilon _{0}c}{\omega ^{2}}(\mathbf{\dot{%
E}\times E)\hat{k}},
\end{equation}%
and they possess SAM with density 
\begin{equation}
\boldsymbol{\rho }_{s}=\frac{\varepsilon _{0}}{\omega ^{2}}(\mathbf{\dot{E}}%
\times \mathbf{E}).
\end{equation}%
The novelty of our work is the method of derivation of the SAM density of
plane waves of light. It is elementary and lengthy, but it is without
reliance on either field theory or quantum mechanics, which makes it
accessible to non-experts in the field. In principle, our calculation could
have been carried out by Maxwell, since it requires only Coulomb's law \cite{depremier} and Maxwell's equations. We believe it clearly shows that plane
waves can carry SAM, quantified by our results as well as Eq.~\eqref{D2}. We
hope that our results, in addition to their pedagogical value, also indicate
the simplicity and usefulness of using the Lorentz force and Maxwell's
equations to describe light - matter interactions.

\nolinenumbers
\bibliographystyle{unsrt}
\bibliography{Torque}

\appendix
\renewcommand{\theequation}{\thesection.\arabic{equation}} 
\setcounter{equation}{0} 

\section{Derivation of the Force Identity}


In this appendix, we provide our derivation of the force identity:%
\begin{equation}
\mathbf{F}_{A}=\int_{0}^{d}\left\langle \mathbf{\dot{P}\times B}%
\right\rangle dz=\frac{1}{c}\left\langle \mathbf{\mathbf{E}}_{i}\mathbf{%
\mathbf{\times H}}_{i}\right\rangle -\frac{1}{c}\left\langle \mathbf{\mathbf{%
E}}_{r}\mathbf{\mathbf{\times H}}_{r}\right\rangle -\frac{1}{c}\left\langle 
\mathbf{\mathbf{\mathbf{E}}}_{t}\mathbf{\mathbf{\mathbf{\times H}}}%
_{t}\right\rangle \mathbf{.}
\end{equation}

\subsection{Macroscopic Fields}

\begin{figure}[]
\centering
\includegraphics[width=0.9\textwidth]{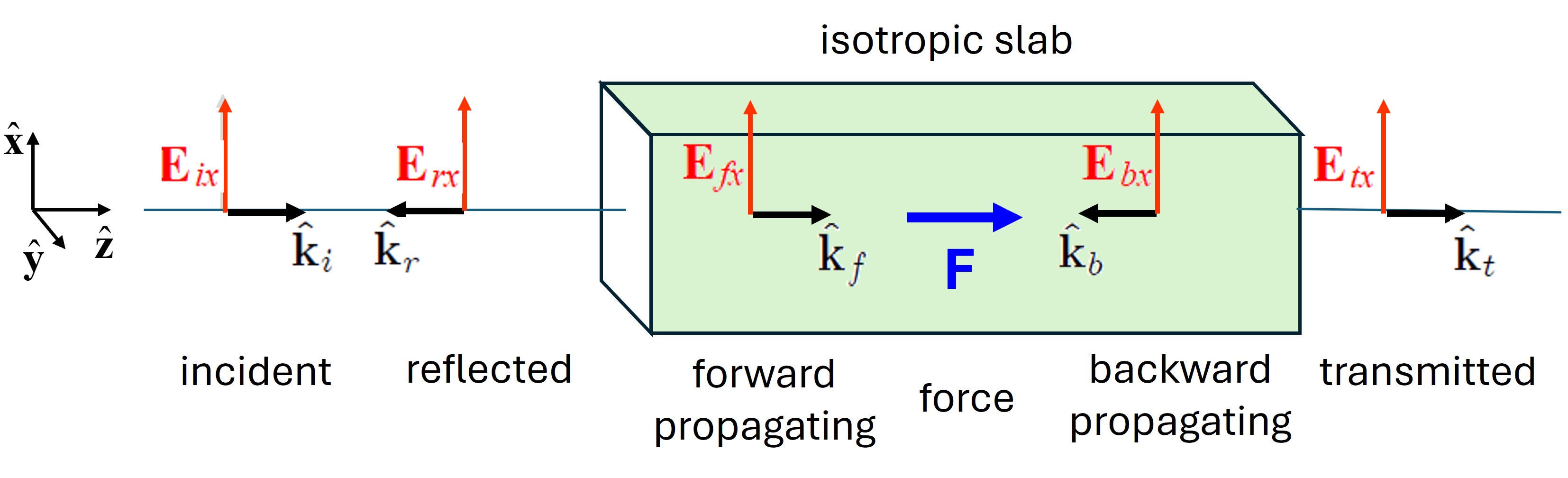}
\caption{Illustration of the sample and fields used in the calculations. The
magnetic fields associated with the electric fields are not shown, they are
in the $\mathbf{\hat{k}\times E}$ direction.}
\label{fig:1}
\end{figure}

We consider an isotropic lossless slab in vacuum, infinite in the $x$ and $y$
directions, with thickness $d$ in the $z$ direction. A schematic is shown in
Fig.~\ref{fig:1}.

The dielectric tensor of the slab can be written as, 
\begin{equation}
\boldsymbol{\varepsilon }=\varepsilon _{0}n^{2}\mathbf{I},
\end{equation}%
where $n$ is the refractive index. The light is normally incident on the
slab in vacuum, and the incident electric field can be expressed as 
\begin{equation}
\mathbf{E}_{i}=E_{i0}\cos (k_{0}z-\omega t+\delta )\mathbf{\hat{x}}.
\end{equation}%
The reflected, transmitted electric field $\mathbf{E}_{r}$, $\mathbf{E}_{t}$%
, and forward and backward electric field in the slab $\mathbf{E}_{f}$, and $%
\mathbf{E}_{b}$ are 
\begin{eqnarray}
\mathbf{E}_{r} &=&E_{i0}A_{1}\cos (-k_{0}z-\omega t+\theta _{1})\mathbf{\hat{%
x},} \\
\mathbf{E}_{f} &=&E_{i0}A_{2}\cos (kz-\omega t+\theta _{2})\mathbf{\hat{x},}
\\
\mathbf{E}_{b} &=&E_{i0}A_{3}\cos (-kz-\omega t+\theta _{3})\mathbf{\hat{x},}
\\
\mathbf{E}_{t} &=&E_{i0}A_{4}\cos (k_{0}(z-d)-\omega t+\theta _{4})\mathbf{%
\hat{x},}
\end{eqnarray}%
and the corresponding magnetic fields are 
\begin{eqnarray}
\mathbf{H}_{r} &=&-\frac{E_{i0}}{Z_{0}}A_{1}\cos (-k_{0}z-\omega t+\theta
_{1})\mathbf{\hat{y},} \\
\mathbf{H}_{f} &=&\frac{E_{i0}}{Z}A_{2}\cos (kz-\omega t+\theta _{2})\mathbf{%
\hat{y},} \\
\mathbf{H}_{b} &=&-\frac{E_{i0}}{Z}A_{3}\cos (-kz-\omega t+\theta _{3})%
\mathbf{\hat{y},} \\
\mathbf{H}_{t} &=&\frac{E_{i0}}{Z_{0}}A_{4}\cos (k_{0}(z-d)-\omega t+\theta
_{4})\mathbf{\hat{y},}
\end{eqnarray}%
where $Z_{0}=\sqrt{\frac{\mu _{0}}{\varepsilon _{0}}}$, $Z=\sqrt{\frac{\mu
_{0}}{\varepsilon }}$, $k_{0}=\frac{2\pi }{\lambda _{0}}$, $k=\frac{2\pi n}{%
\lambda _{0}}$, and the amplitudes and phases are 
\begin{equation}
A=\left( 
\begin{array}{c}
\frac{r\sqrt{2-2\cos 2\phi }}{D} \\ 
\frac{t}{D} \\ 
-\frac{tr}{D} \\ 
\frac{(1-r^{2})}{D}%
\end{array}%
\right) ,  \label{eq_A_F}
\end{equation}%
and%
\begin{equation}
\theta =\left( 
\begin{array}{c}
\beta +\tan ^{-1}(\frac{-\sin 2\phi }{1-\cos 2\phi })+\delta \\ 
\beta +\delta \\ 
\beta +2\phi +\delta \\ 
\beta +\phi +\delta%
\end{array}%
\right) ,  \label{eq_theta_F}
\end{equation}%
where 
\begin{eqnarray}
r &=&\frac{1-n}{1+n}, \\
t &=&\frac{2}{1+n}, \\
\phi &=&\frac{2\pi nd}{\lambda _{0}}, \\
D &=&\sqrt{1-2r^{2}\cos 2\phi +r^{4}}, \\
\beta &=&\tan ^{-1}\left( \frac{r^{2}\sin 2\phi }{1-r^{2}\cos 2\phi }\right)
.
\end{eqnarray}%
We also note that $A_{1}^{2}=4r^{2}\sin ^{2}\phi /D^{2}$, and $%
A_{1}^{2}+A_{4}^{2}=1.$

The free parameters characterizing the system are: $E_{i0}$, $\delta $,$n$,
and $\lambda _{0}$.

\subsection{Force Calculation}

\begin{figure}[]
\centering
\includegraphics[width=0.45\textwidth]{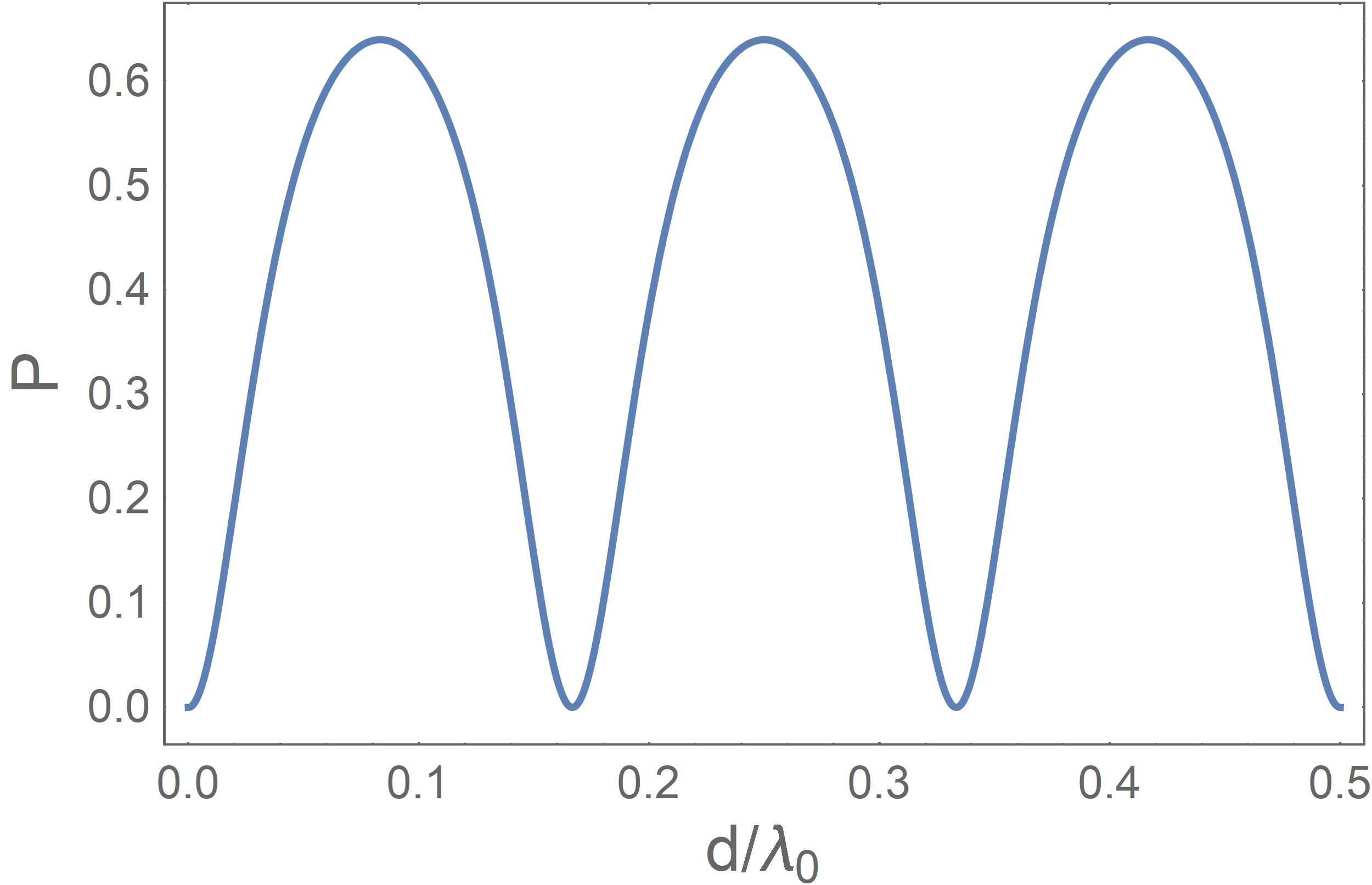} \hfill %
\includegraphics[width=0.45\textwidth]{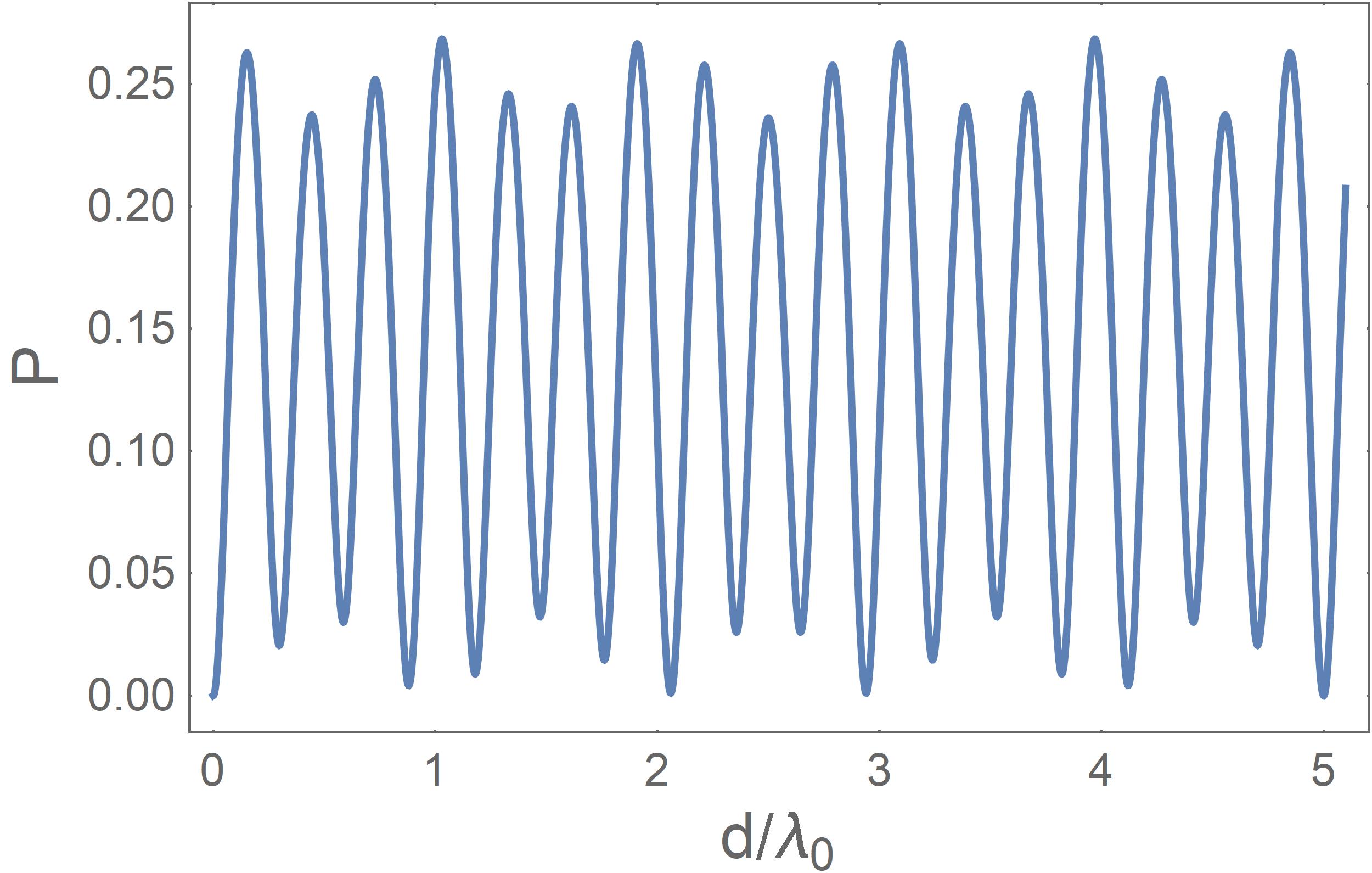}
\caption{Radiation pressure $P$, normalized by $\protect\varepsilon_{0}|%
\mathbf{E}_{i}|^2$, as a function of thickness $d/\protect\lambda_{0}$.
Left: isotropic slab with index $n=3$, described by the Airy function.
Right: waveplate with indices $n_{1}=1.7$ and $n_{2}=1.2$. The radiation pressure $P$ in general is quasiperioedic; here the period is $5$.}
\label{fig:combined}
\end{figure}

Since the electric field inside the slab is the sum of $\mathbf{E}_{f}$ and $%
\mathbf{E}_{b}$, the polarization in the slab is given by 
\begin{equation}
\mathbf{P}=\boldsymbol{\alpha }(\mathbf{E}_{f}+\mathbf{E}_{b}),
\end{equation}%
where the polarizability tensor $\boldsymbol{\alpha }=\boldsymbol{%
\varepsilon }-\varepsilon _{0}\mathbf{I}$. 
\begin{equation}
\mathbf{\dot{P}}\times \mathbf{B}=\varepsilon _{0}(n^{2}-1)(\mathbf{\dot{E}}%
_{f}+\mathbf{\dot{E}}_{b})\times \mu _{0}(\mathbf{H}_{f}+\mathbf{H}_{b}),
\end{equation}%
and after time averaging, we have%
\begin{equation}
\left\langle \mathbf{\dot{P}}\times \mathbf{B}\right\rangle =-\varepsilon
_{0}\mu _{0}\omega \frac{E_{i0}^{2}}{Z}A_{2}A_{3}(n^{2}-1)(\sin (2kz+\theta
_{2}-\theta _{3}))\mathbf{\hat{k}}
\end{equation}%
Integrating over the slab gives%
\begin{eqnarray}
\int_{0}^{d}\left\langle \mathbf{\dot{P}}\times \mathbf{B}\right\rangle dz
&=&-\varepsilon _{0}\mu _{0}\omega \frac{E_{i0}^{2}}{Z}A_{2}A_{3}(n^{2}-1)%
\int_{0}^{d}\sin (2kz+\theta _{2}-\theta _{3}))dz\mathbf{\hat{k}}  \notag \\
&=&\varepsilon _{0}\mu _{0}\omega \frac{E_{i0}^{2}}{Z}A_{2}A_{3}(n^{2}-1)%
\frac{1}{2k}(\cos (2kd+\theta _{2}-\theta _{3})-\cos (\theta _{2}-\theta
_{3}))\mathbf{\hat{k}}  \notag \\
&=&\varepsilon _{0}\mu _{0}\omega \frac{E_{i0}^{2}}{2Zk}%
A_{2}A_{3}(n^{2}-1)(1-\cos 2\phi )\mathbf{\hat{k}}.
\end{eqnarray}%
Substituting for $\theta _{2}$ and $\theta _{3}$ from Eq.~\eqref{eq_theta_F}%
, we get%
\begin{equation}
\int_{0}^{d}\left\langle \mathbf{\dot{P}}\times \mathbf{B}\right\rangle
dz=\varepsilon _{0}\mu _{0}\omega \frac{E_{i0}^{2}}{Zk}A_{2}A_{3}(n^{2}-1)%
\sin ^{2}(kd)\mathbf{\hat{k}.}
\end{equation}%
Noting that 
\begin{eqnarray}
A_{2}A_{3}(n^{2}-1)\sin ^{2}(kd) &=&4\frac{r^{2}}{D^{2}}\sin ^{2}(kd)  \notag
\\
&=&A_{1}^{2}  \notag \\
&=&\frac{1}{2}(1+A_{1}^{2}-A_{4}^{2}),
\end{eqnarray}%
we have%
\begin{equation}
\int_{0}^{d}\left\langle \mathbf{\dot{P}}\times \mathbf{B}\right\rangle dz=%
\frac{\varepsilon _{0}\mu _{0}\omega }{2Zk}(E_{i0}^{2}+E_{r}^{2}-E_{t}^{2})%
\mathbf{\hat{k}},
\end{equation}%
This is our first result. Here we have shown that the force density on the
slab is identically equal to the expression on the right hand side. Noting
that $kZ=k_{0}Z_{0}$ and writing this in covariant form, we obtain our
identity, 
\begin{equation}
\int_{0}^{d}\left\langle \mathbf{\dot{P}}\times \mathbf{B}\right\rangle dz=%
\frac{1}{c}\left\langle \mathbf{\mathbf{E}}_{i}\mathbf{\mathbf{\times H}}%
_{i}\right\rangle -\frac{1}{c}\left\langle \mathbf{\mathbf{\mathbf{E}}}_{r}%
\mathbf{\mathbf{\mathbf{\times H}}}_{r}\right\rangle -\frac{1}{c}%
\left\langle \mathbf{\mathbf{\mathbf{E}}}_{t}\mathbf{\mathbf{\mathbf{\times H%
}}}_{t}\right\rangle .
\end{equation}%
For clarity, we include Fig.~\ref{fig:combined} to illustrate the dependence
of the radiation pressure on sample thickness, or equivalently, on inverse
wavelength.

%

\setcounter{equation}{0} 

\section{Derivation of the Torque Identity}


In this appendix, we provide our derivation of the torque identity:%
\begin{equation}
\mathbf{\tau }_{A}=\int_{0}^{d}\left\langle \mathbf{P\times E}\right\rangle
dz=\frac{\varepsilon _{0}}{k_{0}\omega }(\mathbf{\mathbf{\dot{E}}}_{i}%
\mathbf{\mathbf{\times E}}_{i})\mathbf{-}\frac{\varepsilon _{0}}{k_{0}\omega 
}(\mathbf{\mathbf{\dot{E}}}_{r}\mathbf{\mathbf{\times E}}_{r})\mathbf{-}%
\frac{\varepsilon _{0}}{k_{0}\omega }(\mathbf{\mathbf{\dot{E}}}_{t}\mathbf{%
\mathbf{\times E}}_{t})\mathbf{.}
\end{equation}

\subsection{Macroscopic Fields}

\begin{figure}[b]
\centering
\includegraphics[width=0.9\textwidth]{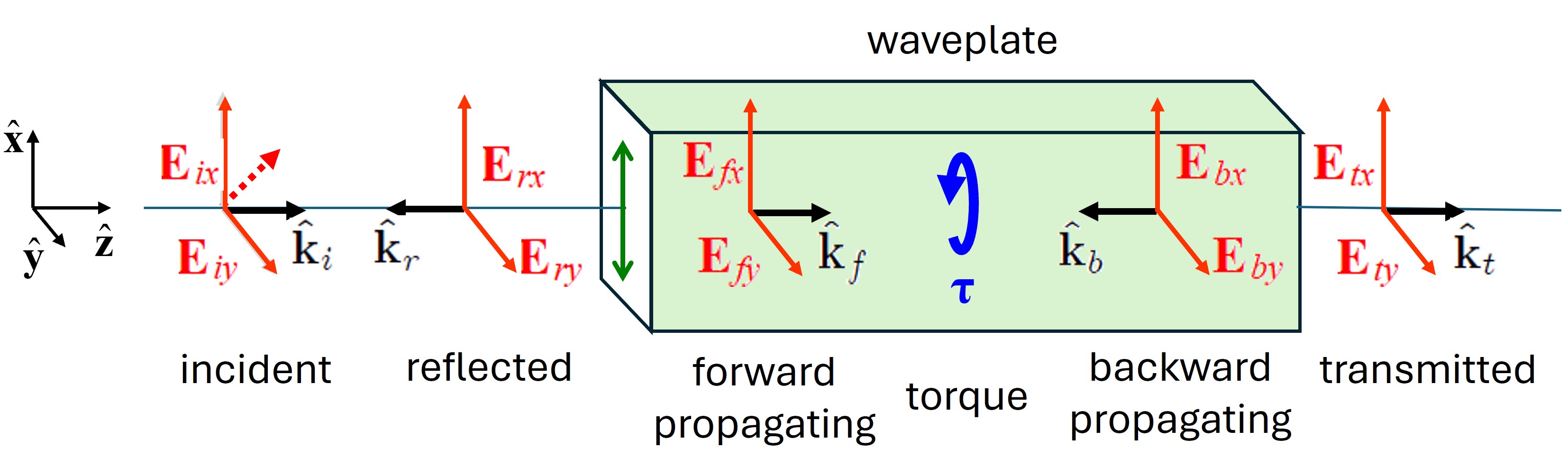}
\caption{Illustration of the sample and fields used in the calculations. The
magnetic fields associated with the electric fields are not shown, they are
in the $\mathbf{\hat{k}\times E}$ direction. The optic axis is indicated
with the double arrow.}
\label{fig:4}
\end{figure}

We consider a uniaxial waveplate in vacuum, infinite in the $x$ and $y$
directions, with thickness $d$ in the $z$ direction; its optic axis is along
the $y$ direction. A schematic is shown in Fig.~\ref{fig:4}.

The dielectric tensor of the waveplate can be written as 
\begin{equation}
\boldsymbol{\varepsilon }=\varepsilon _{0}\left( 
\begin{array}{ccc}
n_{x}^{2} & 0 & 0 \\ 
0 & n_{y}^{2} & 0 \\ 
0 & 0 & n_{z}^{2}%
\end{array}%
\right) ,
\end{equation}%
where $n_{x}=n_{1}$ and $n_{y}=n_{z}=n_{2}$. Light is normally incident on
the waveplate in vacuum.

The incident electric field can be expressed as 
\begin{equation}
\mathbf{E}_{i}=E_{ix0}\cos (k_{0}z-\omega t+\delta _{1})\mathbf{\hat{x}}%
+E_{iy0}\cos (k_{0}z-\omega t+\delta _{2})\mathbf{\hat{y}},
\end{equation}%
then the reflected, transmitted electric fields $\mathbf{E}_{r}$, $\mathbf{E}%
_{t}$, and forward and backward electric fields $\mathbf{E}_{f}$, and $%
\mathbf{E}_{b}$ in the waveplate are 
\begin{eqnarray}
\mathbf{E}_{r} &=&E_{ix0}A_{1,1}\cos (-k_{0}z-\omega t+\theta _{1,1})\mathbf{%
\ \hat{x}}+E_{i0y}A_{1,2}\cos (-k_{0}z-\omega t+\theta _{1,2})\mathbf{\hat{y}%
}, \\
\mathbf{E}_{f} &=&E_{ix0}A_{2,1}\cos (k_{1}z-\omega t+\theta _{2,1})\mathbf{%
\ \hat{x}}+E_{iy0}A_{2,2}\cos (k_{2}z-\omega t+\theta _{2,2})\mathbf{\hat{y}}%
, \\
\mathbf{E}_{b} &=&E_{ix0}A_{3,1}\cos (-k_{1}z-\omega t+\theta _{3,1})\mathbf{%
\ \hat{x}}+E_{iy0}A_{3,2}\cos (-k_{2}z-\omega t+\theta _{3,2})\mathbf{\hat{y}%
}, \\
\mathbf{E}_{t} &=&E_{ix0}A_{4,1}\cos (k_{0}(z-d)-\omega t+\theta _{4,1})%
\mathbf{\hat{x}}+E_{iy0}A_{4,2}\cos (k_{0}(z-d)-\omega t+\theta _{4,2})%
\mathbf{\hat{y}},
\end{eqnarray}%
where $k_{0}=\frac{2\pi }{\lambda _{0}}$, $k_{1}=\frac{2\pi n_{1}}{\lambda
_{0}}$, $k_{2}=\frac{2\pi n_{2}}{\lambda _{0}}$, and the amplitudes and
phases are 
\begin{equation}
A=\left( 
\begin{array}{cc}
\frac{r_{1}\sqrt{2-2\cos 2\phi _{1}}}{D_{1}} & \frac{r_{2}\sqrt{2-2\cos
2\phi _{2}}}{D_{2}} \\ 
\frac{t_{1}}{D_{1}} & \frac{t_{2}}{D_{2}} \\ 
-\frac{t_{1}r_{1}}{D_{1}} & -\frac{t_{2}r_{2}}{D_{2}} \\ 
\frac{(1-r_{1}^{2})}{D_{1}} & \frac{(1-r_{2}^{2})}{D_{2}}%
\end{array}%
\right) ,  \label{eq_A}
\end{equation}%
and%
\begin{equation}
\theta =\left( 
\begin{array}{cc}
\beta _{1}+\tan ^{-1}(\frac{-\sin 2\phi _{1}}{1-\cos 2\phi _{1}})+\delta _{1}
& \beta _{2}+\tan ^{-1}(\frac{-\sin 2\phi _{2}}{1-\cos 2\phi _{2}})+\delta
_{2} \\ 
\beta _{1}+\delta _{1} & \beta _{2}+\delta _{2} \\ 
\beta _{1}+2\phi _{1}+\delta _{1} & \beta _{2}+2\phi _{2}+\delta _{2} \\ 
\beta _{1}+\phi _{1}+\delta _{1} & \beta _{2}+\phi _{2}+\delta _{2}%
\end{array}%
\right) ,  \label{eq_theta}
\end{equation}%
where 
\begin{eqnarray}
r_{1} &=&\frac{1-n_{1}}{1+n_{1}},~r_{2}=\frac{1-n_{2}}{1+n_{2}}, \\
t_{1} &=&\frac{2}{1+n_{1}},~t_{2}=\frac{2}{1+n_{2}}, \\
\phi _{1} &=&\frac{2\pi n_{1}d}{\lambda _{0}},~\phi _{2}=\frac{2\pi n_{2}d}{%
\lambda _{0}}, \\
D_{1} &=&\sqrt{1-2r_{1}^{2}\cos 2\phi _{1}+r_{1}^{4}},~D_{2}=\sqrt{%
1-2r_{2}^{2}\cos 2\phi _{2}+r_{2}^{4}}, \\
\beta _{1} &=&\tan ^{-1}\left( \frac{r_{1}^{2}\sin 2\phi _{1}}{%
1-r_{1}^{2}\cos 2\phi _{1}}\right) ,~\beta _{2}=\tan ^{-1}\left( \frac{%
r_{2}^{2}\sin 2\phi _{2}}{1-r_{2}^{2}\cos 2\phi _{2}}\right) .
\end{eqnarray}

The free parameters characterizing the system are: $E_{ix0}$, $\delta _{1}$, 
$E_{iy0}$, $\delta _{2}$, $n_{1}$, $n_{2}$ and $\lambda _{0}$.

We remark that at this point, sufficient information has been provided to
verify our identity numerically.

\subsection{Torque Calculation}

The polarization in the waveplate is given by 
\begin{equation}
\mathbf{P}=\boldsymbol{\alpha }(\mathbf{E}_{f}+\mathbf{E}_{b}),
\end{equation}%
where the polarizability tensor $\boldsymbol{\alpha }=\boldsymbol{\
\varepsilon }-\varepsilon _{0}\mathbf{I}$. 
\begin{eqnarray}
\mathbf{P}\times (\mathbf{E}_{f}+\mathbf{E}_{b}) &=&(\alpha _{xx}\mathbf{\ 
\hat{x}\hat{x}}+\alpha _{yy}\mathbf{\hat{y}\hat{y})}(\mathbf{E}_{f}+\mathbf{E%
}_{b})\times (\mathbf{E}_{f}+\mathbf{E}_{b})  \notag \\
&=&(\alpha _{xx}-\alpha _{yy})(\mathbf{\hat{x}\cdot (E}_{f}+\mathbf{E}_{b}))(%
\mathbf{\hat{y}}\cdot (\mathbf{E}_{f}+\mathbf{E}_{b}))\mathbf{\hat{z}}, 
\notag \\
&=&\varepsilon _{0}\mathbf{(}n_{1}^{2}-n_{2}^{2}\mathbf{)}E_{ix0}E_{iy0}\cdot
\notag \\
&&(A_{2,1}\cos (k_{1}z-\omega t+\theta _{2,1})+A_{3,1}\cos (-k_{1}z-\omega
t+\theta _{3,1}))\cdot  \notag \\
&&(A_{2,2}\cos (k_{2}z-\omega t+\theta _{2,2})+A_{3,2}\cos (-k_{2}z-\omega
t+\theta _{3,2}))\mathbf{\hat{z}}.
\end{eqnarray}%
After integrating, averaging over time and defining%
\begin{equation}
\tau _{0}=\frac{\varepsilon _{0}E_{ix0}E_{iy0}\lambda _{0}}{2\pi },
\end{equation}%
the dimensionless areal torque density $\tau _{A}/\tau _{0}$ is given by 
\begin{eqnarray}
\tau _{A}/\tau _{0} &=&2\eta _{a}A_{2,1}A_{2,2}\cos (\theta _{2,1}-\theta
_{2,2}+\frac{1}{2}(\phi _{1}-\phi _{2}))\sin (\frac{1}{2}(\phi _{1}-\phi
_{2}))+  \notag  \label{eq_tau} \\
&&2\eta _{d}A_{2,1}A_{3,2}\cos (\theta _{2,1}-\theta _{3,2}+\frac{1}{2}(\phi
_{1}+\phi _{2}))\sin (\frac{1}{2}(\phi _{1}+\phi _{2}))+  \notag \\
&&2\eta _{d}A_{3,1}A_{2,2}\cos (\theta _{3,1}-\theta _{2,2}-\frac{1}{2}(\phi
_{1}+\phi _{2}))\sin (\frac{1}{2}(\phi _{1}+\phi _{2}))+  \notag \\
&&2\eta _{a}A_{3,1}A_{3,2}\cos (\theta _{3,1}-\theta _{3,2}-\frac{1}{2}(\phi
_{1}-\phi _{2}))\sin (\frac{1}{2}(\phi _{1}-\phi _{2})),
\end{eqnarray}%
where%
\begin{equation}
\eta _{a}=\frac{1}{2}(n_{1}+n_{2}),~\eta _{d}=\frac{1}{2}(n_{1}-n_{2}).
\end{equation}%
We next define 
\begin{equation}
B=D_{1}D_{2}\tau _{A}/\tau _{0}.
\end{equation}%
The quantity $B$ is essentially the right hand side of Eq.~\eqref{eq_tau},
multiplied by the factor $D_{1}D_{2}$. The majority of remaining effort is
to simplify the expression for $B$ via algebra and trigonometry identities.

We define the differences 
\begin{eqnarray}
dd &=&\delta _{1}-\delta _{2}, \\
df &=&\phi _{1}-\phi _{2}, \\
db &=&\beta _{1}-\beta _{2},
\end{eqnarray}%
and then can write 
\begin{eqnarray}
B &=&t_{1}t_{2}((1-r_{1}r_{2})\eta _{a}\sin (db+dd+df)-\eta _{a}\sin
(db+dd)+(r_{1}-r_{2})\eta _{d}\sin (db+dd+df)  \notag \\
&&+r_{2}\eta _{d}\sin (db+dd-2\phi _{2})-r_{1}\eta _{d}\sin (db+dd+2\phi
_{1})+r_{1}r_{2}\eta _{a}\sin (db+dd+2df)).
\end{eqnarray}

To facilitate simplification, we write%
\begin{equation}
B=B_{1}+B_{2},
\end{equation}%
where%
\begin{equation}
B_{1}=t_{1}t_{2}((1-r_{1}r_{2})\eta _{a}\sin (db+dd+df)+(r_{1}-r_{2})\eta
_{d}\sin (db+dd+df)),  \label{eq_B1}
\end{equation}%
and 
\begin{equation}
B_{2}=t_{1}t_{2}(-\eta _{a}\sin (db+dd)+r_{2}\eta _{d}\sin (db+dd-2\phi
_{2})-r_{1}\eta _{d}\sin (db+dd+2\phi _{1})+r_{1}r_{2}\eta _{a}\sin
(db+dd+2df)).  \label{eq_B2}
\end{equation}%
Noting that%
\begin{equation}
t_{1}t_{2}((1-r_{1}r_{2})\eta _{a}+(r_{1}-r_{2})\eta
_{d})=(1-r_{1}^{2})(1-r_{2}^{2}),
\end{equation}%
substituting into Eq.~\eqref{eq_B1}, $B_{1}$ becomes 
\begin{equation}
B_{1}=(1-r_{1}^{2})(1-r_{2}^{2})\sin (db+df+dd).
\end{equation}%
Noting that%
\begin{equation}
t_{1}t_{2}\eta _{a}=1-r_{1}r_{2},
\end{equation}%
and%
\begin{equation}
t_{1}t_{2}\eta _{d}=r_{2}-r_{1},
\end{equation}%
with substitution into Eq.~\eqref{eq_B2}, $B_{2}$ becomes 
\begin{eqnarray}
B_{2} &=&r_{1}r_{2}(\sin (db+dd)-\sin (db+dd-2\phi _{2})-\sin (db+dd+2\phi
_{1})+\sin (db+dd+2df))  \notag  \label{eq_43} \\
&-&\sin (db+dd)+r_{2}^{2}\sin (db+dd-2\phi _{2})+r_{1}^{2}\sin (db+dd+2\phi
_{1})-r_{1}^{2}r_{2}^{2}\sin (db+dd+2df).
\end{eqnarray}%
We next write 
\begin{equation}
B_{2}=r_{1}r_{2}(\sin (db+dd)-\sin (db+dd-2\phi _{2})-\sin (db+dd+2\phi
_{1})+\sin (db+dd+2df))+B_{3},
\end{equation}%
where we have defined $B_{3}$ as the expressions in the second line of Eq.~%
\eqref{eq_43}.

Noting that 
\begin{equation}
D_{1}D_{2}\cos (db)=1-r_{1}^{2}\cos (2\phi _{1})-r_{2}^{2}\cos (2\phi
_{2})+r_{1}^{2}r_{2}^{2}\cos (2df),
\end{equation}%
and%
\begin{equation}
D_{1}D_{2}\sin (db)=r_{1}^{2}\sin (2\phi _{1})-r_{2}^{2}\sin (2\phi
_{2})-r_{1}^{2}r_{2}^{2}\sin (2df),
\end{equation}%
we evaluate and simplify 
\begin{eqnarray}
B_{3} &=&-\sin (db+dd)+r_{2}^{2}\sin (db+dd-2\phi _{2})+r_{1}^{2}\sin
(db+dd+2\phi _{1})-r_{1}^{2}r_{2}^{2}\sin (db+dd+2df)  \notag \\
&=&-\sin (db+dd)+r_{2}^{2}\sin (db+dd)\cos (2\phi _{2})-r_{2}^{2}\cos
(db+dd)\sin (2\phi _{2})  \notag \\
&&+r_{1}^{2}\sin (db+dd)\cos (2\phi _{1})+r_{1}^{2}\cos (db+dd)\sin (2\phi
_{1})  \notag \\
&&-r_{1}^{2}r_{2}^{2}\sin (db+dd)\cos (2df)-r_{1}^{2}r_{2}^{2}\cos
(db+dd)\sin (2df)  \notag \\
&=&-\sin (db+dd)(1-r_{2}^{2}\cos (2\phi _{2})-r_{1}^{2}\cos (2\phi
_{1})+r_{1}^{2}r_{2}^{2}\cos (2df))  \notag \\
&&+\cos (db+dd)(-r_{2}^{2}\sin (2\phi _{2})+r_{1}^{2}\sin (2\phi
_{1})-r_{1}^{2}r_{2}^{2}\sin (2df))  \notag \\
&=&-D_{1}D_{2}\sin (db+dd)\cos (db)+D_{1}D_{2}cos(db+dd)\sin (db)  \notag \\
&=&-D_{1}D_{2}\sin (dd).
\end{eqnarray}%
Then we have%
\begin{eqnarray}
B_{2} &=&r_{1}r_{2}(\sin (db+dd)-\sin (db+dd-2\phi _{2})-\sin (db+dd+2\phi
_{1})+\sin (db+dd+2df))-D_{1}D_{2}\sin (dd)  \notag \\
&=&-D_{1}D_{2}\sin (dd)+4r_{1}r_{2}\sin (\phi _{1})\sin (\phi _{2})\sin
(db+df+dd),
\end{eqnarray}%
and finally, together with Eq.~\eqref{eq_B1}, we have 
\begin{equation}
B=-D_{1}D_{2}\sin (dd)+4r_{1}r_{2}\sin (\phi _{1})\sin (\phi _{2})\sin
(db+df+dd)+(1-r_{1}^{2})(1-r_{2}^{2})\sin (db+df+dd).
\end{equation}%
Then, since $\tau _{A}/\tau _{0}=B/(D_{1}D_{2})$, we have for the
dimensionless areal torque density 
\begin{eqnarray}
\tau _{A}/\tau _{0} &=&-\sin (dd)+4\frac{r_{1}r_{2}}{D_{1}D_{2}}\sin (\phi
_{1})\sin (\phi _{2})\sin (db+df+dd)+\frac{(1-r_{1}^{2})(1-r_{2}^{2})}{%
D_{1}D_{2}}\sin (db+df+dd)  \notag \\
&=&-\sin (\delta _{1}-\delta _{2})+A_{1,1}A_{1,2}\sin (\theta _{1,1}-\theta
_{1,2})+A_{4,1}A_{4,2}\sin (\theta _{4,1}-\theta _{4,2}).
\end{eqnarray}%
Returning to dimensional units, we have%
\begin{equation}
\tau _{A}=-\frac{\varepsilon _{0}\lambda _{0}}{2\pi }(E_{ix0}E_{iy0}\sin
(\delta _{{i}x}-\delta _{{i}y}) - E_{rx0}E_{ry0}\sin (\delta _{rx}-\delta
_{ry}) - E_{tx0}E_{ty0}\sin (\delta _{tx}-\delta _{ty})),
\end{equation}%
or 
\begin{equation}
\tau _{A}=-\frac{\varepsilon _{0}c}{\omega }(E_{ix0}E_{iy0}\sin (\delta _{{i}%
x}-\delta _{{i}y}) - E_{rx0}E_{ry0}\sin (\delta _{rx}-\delta _{ry}) -
E_{tx0}E_{ty0}\sin (\delta _{tx}-\delta _{ty})),
\end{equation}%
where $\delta _{rx}-\delta _{ry}=\theta _{1,1}-\theta _{1,2}\,\ $and $\delta
_{tx}-\delta _{ty}=\theta _{4,1}-\theta _{4,2}$.

In summary, we have shown that the areal torque density on the waveplate is
identically equal to the expression on the right hand side. %
Writing this in covariant form, we obtain our identity 
\begin{equation}
\mathbf{\tau }_{A}=\int_{0}^{d}\left\langle \mathbf{P\times E}\right\rangle
dz=\frac{\varepsilon _{0}c}{\omega ^{2}}(\mathbf{\mathbf{\dot{E}}}_{i}%
\mathbf{\mathbf{\times E}}_{i})\mathbf{-}\frac{\varepsilon _{0}c}{\omega ^{2}%
}(\mathbf{\mathbf{\dot{E}}}_{r}\mathbf{\mathbf{\times E}}_{r})\mathbf{-}%
\frac{\varepsilon _{0}c}{\omega ^{2}}(\mathbf{\mathbf{\dot{E}}}_{t}\mathbf{%
\mathbf{\times E}}_{t}).
\end{equation}%
This is our second and main result.

\begin{figure}[tbp]
\centering
\includegraphics[width=1\textwidth]{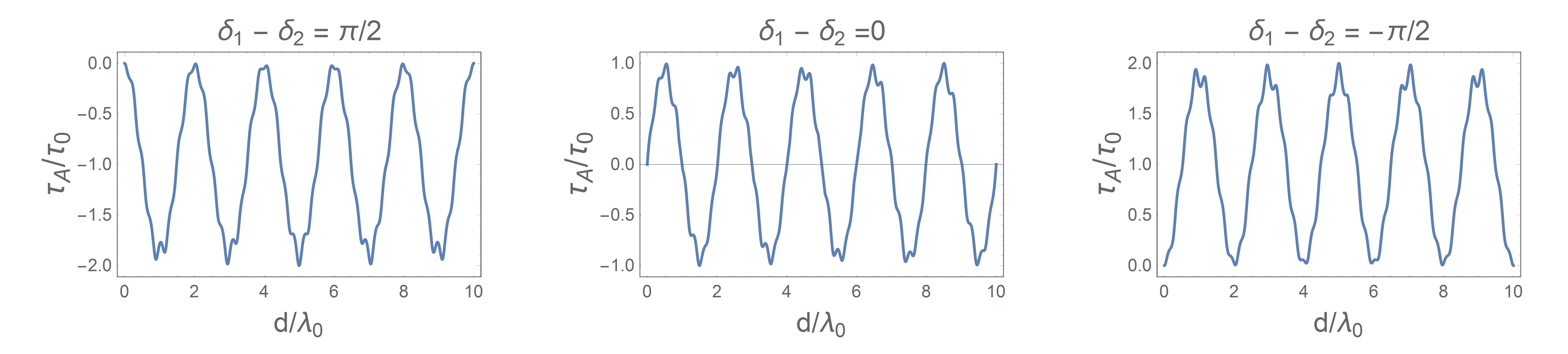}
\caption{Areal torque density $\mathbf{\protect\tau }_{A}/\protect\tau _{0}$%
, on a wave plate with indices $n_{1}=1.7$ and $n_{2}=1.2$ as function of
thickness $d/\protect\lambda _{0}$ for (left) left-circular, (middle) linear
and (right) right-circular polarizations, from the point of view of the
source. The areal torque density in general is quasiperiodic; here the
period is $10$.}
\label{fig:5}
\end{figure}

For clarity, we include Fig.~\ref{fig:5} to indicate the dependence of the
areal torque density of sample thickness, or equivalently, inverse
wavelength. We also show the effect of polarization of the incident light.

\end{document}